\documentclass[sn-basic]{sn-jnl}% Basic Springer Nature Reference Style/Chemistry Reference Style

\usepackage{graphicx}%
\usepackage{multirow}%
\usepackage{amsmath,amssymb,amsfonts}%
\usepackage{amsthm}%
\usepackage{mathrsfs}%
\usepackage[title]{appendix}%
\usepackage{xcolor}%
\usepackage{textcomp}%
\usepackage{manyfoot}%
\usepackage{booktabs}%
\usepackage{algorithm}%
\usepackage{algorithmicx}%
\usepackage{algpseudocode}%
\usepackage{listings}%
\usepackage{xcolor}%
\usepackage{lmodern}
\begin{document}

% Title
\title[Continuing to Advance European High-Contrast Imaging R\&D towards HWO and LIFE]{Continuing to Advance European High-Contrast Imaging R\&D towards HWO and LIFE}

% Authors
\author*[1,2]{\fnm{Ga\"el} \sur{Chauvin}}\email{chauvin@mpia.de}

\author[1]{\fnm{Óscar} \sur{Carrión-González}}\email{oscarrion@mpia.de}

\author[2,3]{\fnm{Iva} \sur{Laginja}}\email{iva.laginja@obspm.fr}

\author[4]{\fnm{Daniel} \sur{Dicken}}\email{daniel.dicken@stfc.ac.uk <daniel.dicken@stfc.ac.uk}

% WFC & S
\author[5]{\fnm{Sebastiaan} \sur{Haffert}}\email{haffert@strw.leidenuniv.nl}

\author[6]{\fnm{Markus} \sur{Kasper}}\email{kasper@eso.org}

% Ground-based
\author[7]{\fnm{Olivier} \sur{Absil}}\email{olivier.absil@uliege.be}

\author[6]{\fnm{Jens} \sur{Kammerer}}\email{jkammere@eso.org}

% Corono
\author[3]{\fnm{Axel} \sur{Potier}}\email{axel.potier@obspm.fr}

% Data processing
\author[8]{\fnm{Hervé} \sur{Le Coroller}}\email{hereve.lecoroller@lam.fr}

\author[1]{\fnm{Elisabeth} \sur{Matthews}}\email{matthews@mpia.de}

% Nulling
\author[9]{\fnm{Romain} \sur{Laugier}}\email{romain.laugier@kuleuven.be}

\author[9]{\fnm{Denis} \sur{Defrère}}\email{denis.defrere@kuleuven.be}

\author[10]{\fnm{Jonah} \sur{Hansen}}\email{johansen@phys.ethz.ch}

% Telescope
\author[1]{\fnm{Oliver} \sur{Krause}}\email{krause@mpia.de}

% Detectors
\author[11]{\fnm{Pieter} \sur{de Visser}}\email{p.j.de.visser@sron.nl}

\author[12]{\fnm{Mikael} \sur{Karlsson}}\email{mikael.karlsson@angstrom.uu.se}

\author[1]{\fnm{Thomas} \sur{Henning}}\email{henning@mpia.de}

\author[8]{\fnm{Marc} \sur{Ferrari}}\email{marc.ferrari@osupytheas.fr}

\author[13]{\fnm{Jonas} \sur{Kuehn}}\email{jonas.kuehn@unibe.ch}

\author[14]{\fnm{Markus} \sur{Janson}}\email{markus.janson@astro.su.se}

\author[15]{\fnm{Feng} \sur{Zhao}}\email{Feng.Zhao@jpl.nasa.gov}

\author[16]{\fnm{C\'elia} \sur{Desgrange}}\email{Celia.Desgrange@eso.org}

\affil[1]{\orgdiv{MPIA}, \orgname{Max-Planck-Institut für Astronomie}, \orgaddress{Königstuhl 17}, \city{Heidelberg}, \postcode{D-69117}, \country{Germany}}

\affil[2]{Laboratoire J.-L. Lagrange, Université Cote d’Azur, CNRS, Observatoire de la Cote d’Azur, 06304 Nice, France}

\affil[3]{\orgdiv{LIRA, Observatoire de Paris}, \orgname{Université PSL, CNRS, Sorbonne Université, Université Paris Cité}, \orgaddress{\street{5 place Jules Janssen}, \city{Meudon}, \postcode{92195}, \country{France}}}

\affil[4]{UK Astronomy Technology Centre, Royal Observatory, Blackford
Hill Edinburgh, EH9 3HJ Scotland, UK} 

\affil[5]{\orgdiv{NOVA}, \orgname{Leiden University}, \orgaddress{\street{Einsteinweg 55}, \city{Leiden}, \postcode{2333 CC}, \country{The Netherlands}}}

\affil[6]{European Southern Observatory, Karl-Schwarzschild-Strasse 2, 85748 Garching bei München, Germany}

\affil[7]{\orgdiv{Space sciences, Technologies, and Astrophysics Research (STAR) Institute}, \orgname{Universit\'e de Li\`ege}, \orgaddress{all\'ee du Six Ao\^ut 19c}, \city{Li\`ege}, \postcode{4000}, \country{Belgium}}

%\affil[9]{\orgdiv{LIRA, Observatoire de Paris}, \orgname{Université PSL, CNRS, Sorbonne Université, Université Paris Cité}, \orgaddress{\street{5 place Jules Janssen}, \city{Meudon}, \postcode{92195}, \country{France}}}

\affil[8]{Aix Marseille Univ, CNRS, CNES, LAM, Marseille, France}

\affil[9]{Institute of Astronomy, KU Leuven, Celestijnenlaan 200D, 3001, Leuven, Belgium}

\affil[10]{ETH Zurich, Institute for Particle Physics and Astrophysics, Wolfgang-Pauli-Str. 27, 8093 Zurich, Switzerland}

\affil[11]{SRON - Space Research Organisation Netherlands, Niels Bohrweg 4, 2333 CA Leiden, The Netherlands}

%\affil[13]{ETH Zurich, Institute for Particle Physics and Astrophysics, Wolfgang-Pauli-Str. 27, 8093 Zurich, Switzerland}

\affil[12]{Department of Materials Science and Engineering, Uppsala
University, P.O. Box 35, Uppsala, 751 03, Sweden}

\affil[13]{Division of Space and Planetary Sciences, University of Bern, Sidlerstrasse 5, 3012 Bern,
Switzerland}

\affil[14]{Department of Astronomy, Stockholm University, AlbaNova University Center, 10691 Stockholm, Sweden}

\affil[15]{Jet Propulsion Laboratory, California Institute of Technology, 4800 Oak Grove Drive, Pasadena CA 91109, USA}

\affil[16]{European Southern Observatory, Alonso de Córdova 3107, Vitacura, Santiago, Chile}

\abstract{The European R\&D for Space-based High-Contrast Imaging (HCI) II Workshop, held at MPIA in May 2025, advanced Europe’s strategic coordination in support of future exoplanet-imaging missions such as the Habitable Worlds Observatory (HWO) and the Large Interferometer for Exoplanets (LIFE) mission. Building on the first 2024 workshop, this meeting defined concrete priorities across eight technical areas, including wavefront sensing, coronagraphs, post-processing, nulling interferometry, deformable mirrors, detectors, and telescope design. Discussions emphasized Europe’s strengths in adaptive optics, ground-based facilities, and interferometry, while identifying key gaps, particularly the need for a dedicated European vacuum testbed for high-contrast imaging. The community highlighted near-infrared or UV coronagraphy as a promising domain for European leadership and called for joint development of advanced data reduction algorithms,  detectors, and cross-mission coordination with HWO and LIFE. The workshop outcomes establish a collaborative roadmap to strengthen Europe’s technological readiness, foster agency partnerships, and ensure its continued leadership in the next generation of space-based exoplanet exploration.}




\maketitle

%\tableofcontents % temporary

% Introduction
\section{Introduction}

% - HWO/LIFE/Roman context \& timeline\\
% - ESA's Voyage 2050 report and US Decadal Survey identified HCI as a priority; the EU community gathered to move forward with that
% - previous workshop, previous workshop conclusions (Iva's intro talk and slides)\\
% - NASA people were present at this meeting, and (not official, but scientific) ESA reps\\
% - setting the stage with this year's workshop being more concrete since we went into concrete technical HCI topic discussions; develop the objectives the workshop and the identified priorities and work packages with regards to expertise in Europe\\
% - timeliness of this due to what's going on right now: HWO project office was established - June 2024, first official HWO science conference July 2025, ESA kinda taking steps forward as announced in their HWO town hall - May 2025, LIFE upcoming conference  - November 20025, ESA ministerial - November 2025 \\
% - it's a community report, milestone for us (the HCI community), keep talking to agencies and institutions\\

The search for life beyond Earth has long been a central theme in astronomy, with missions such as the Large Interferometer For Exoplanets \citep[LIFE,][]{quanzetal2022} and the  Habitable Worlds Observatory \citep[HWO,][]{Feinberg2026} representing major steps forward in this endeavour. These initiatives are situated within a broader historical and scientific context, including the Nancy Grace Roman Space Telescope \citep{spergeletal2015, baileyetal2023} and other flagship missions, which together form a timeline of increasingly ambitious efforts to detect and characterize habitable-zone exoplanets.

Both the European Space Agency (ESA) and the National Aeronautics and Space Administration (NASA) have recognized the importance of high-contrast imaging (HCI) for future exoplanet missions. ESA's \textit{Voyage 2050} report\footnote{\url{https://www.cosmos.esa.int/web/voyage-2050}} and the US Decadal Survey \citep{astro2020} have independently identified HCI as a strategic priority. In response, the European scientific community has begun to organize and mobilize around this goal, building momentum through collaborative workshops and strategic planning. A key milestone in this process was the first R\&D for Space-based HCI workshop organized in May 2024 in Paris\footnote{https://hcieurope.sciencesconf.org/} (hereafter RD-HCI I), whose conclusions are summarized in \citet{2025Ap&SS.370...29L}
and helped defining the scientific and technical landscape. 

In May 2025, the second edition of the R\&D for Space-based HCI workshop was organized at MPIA/Heidelberg\footnote{https://hcieurope-mpia.sciencesconf.org/} (hereafter RD-HCI II) and marked a significant evolution in focus and ambition. Moving beyond general discussions, the meeting concentrated on concrete technical topics in HCI, aiming to define clear objectives, identify priority areas, and develop work packages aligned with European expertise. This shift reflects both the maturity of the community and the urgency of the moment. The timeliness of this effort is underscored by several recent developments: the establishment of the HWO project office in June 2024, the first official HWO science conference held in July 2025, ESA's announcements during the HWO town hall in May 2025, the LIFE conference in November 2025 and the ESA Ministerial Council meeting scheduled for the same month -- which resulted in the largest member states contribution in the history of ESA, with €22.3bn\footnote{\url{https://www.esa.int/Newsroom/Press_Releases/ESA_Member_States_commit_to_largest_contributions_at_Ministerial}}. These events signal a period of rapid progress and decision-making, making it essential for the European HCI community to articulate its vision and capabilities.

This report represents a milestone for the European HCI community. It is a collective effort to consolidate expertise, define strategic directions, and maintain active dialogue with space agencies and institutions. By doing so, we aim to ensure that Europe remains at the forefront of the next generation of exoplanet exploration.

As part of the RD-HCI II organization, following updates on the Roman, HWO, and LIFE projects, as well as R\&D activities at ESO, ESA, and within national communities across Europe, the workshop transitioned into focused breakout discussions on specific topical work packages that had been identified and prioritized by participants before the workshop. The eight work packages included: WP1 – Wavefront Sensing, Control \& AO; WP2 – Ground-Based Facilities; WP3 – Coronagraphs; WP4 – Post-Processing; WP5 – Nulling; WP6 – Deformable Mirrors; WP7 – Telescope Design \& Optics; and WP8 – Detectors. The final part of the workshop was devoted to presenting the main conclusions from these discussions in plenary sessions and defining key actions for the coming months and years. The following sections summarize the content of these discussions and the principal outcomes of the breakout sessions.

\section{Wavefront Sensing, Control \& Adaptive Optics}
% Sebastiaan Haffert
Wavefront sensing and control is the cornerstone of the coronagraphic instrument for the HWO mission. Wavefront disturbances caused by imperfect optics and telescope deformation will cause light to leak through the coronagraph and overwhelm the planet signal. Many institutes within Europe have a strong heritage in adaptive optics and wavefront sensing and control \citep{ragazzoni2000testing, rousset2003naos, haffert2018sky, beuzit2019sphere, fetick2023papyrus, allouche2025first}. Most of the wavefront sensors that are currently in use or are planned were originally developed in Europe (PyWFS: \cite{ragazzoni1996pupil,feldt2006}, ZWFS: \cite{zernike1935phase}). This development has primarily occurred in ground-based environments. However, the environment, requirements, and goals for coronagraphic instruments are different. This means that there is still a lot of work to do to adapt to space-based conditions.\vspace{0.1cm}

\textbf{Advanced control algorithms} Most of the current wavefront sensing and control algorithms are linearized. There is room for substantial improvement because most wavefront sensors are inherently non-linear \citep{burvall2006linearity, n2013calibration}. Much of ground-based wavefront sensing and control research focuses on machine learning. Little work has been done so far on applying machine learning to space-based HCI problems, although new developments are being fostered by JWST \citep[e.g.][]{desdoigtsetal2025arXiv251009806D}.\vspace{0.1cm}

\textbf{Optimal electric field sensing techniques} There are two electric field sensing techniques at the moment for HCI: pair-wise probing (PWP) and the Self-Coherent Camera (SCC). These are well developed both in their modeling effort and with lab-based verification. However, they are not necessarily optimal \citep{laginja2025extended}. Advances in non-linear optimization through machine learning may enable novel wavefront sensing and control architectures that outperform current methods \citep[e.g.,][]{OrbandeXivry2021,Gutierrez2024,Nousiainen2026}.\vspace{0.1cm}

\textbf{Real-time computing} Many real-time computing systems have been developed within Europe for adaptive optics, and we have a strong heritage in this technology \citep{bitenc2015durham,ferreira2020hard,thiebaut2022closing,fetick2023papyrus}. However, we have not yet explored this for space-based AO systems. Adapting our current systems to space environments will require testing many different aspects such as radiation tolerance. Additionally, for space missions, other aspects are crucial such as redundancy and memory size. Furthermore, most of the software libraries are lagging behind. The space sector is quite risk-averse, and it is difficult to fly commercial processors and software.\vspace{0.1cm}

\textbf{Testbed development} High-contrast imaging models and simulations are theoretically well developed. However, the transition from modeling to lab experiments shows that at the required wavefront stability (at the picometer level), many components do not behave as assumed. A crucial step is the development of end-to-end verification with real instruments. However, there are currently no vacuum testbeds available in Europe for high-contrast imaging purposes. Collaborating with vacuum testbeds at NASA's Jet Propulsion Laboratory (JPL), where multiple testbeds have been developed, is a very bureaucratic process for external experiments. Experience shows that there is a long cycle from idea to implementation that can be up to multiple years. This is too long for the targeted cycle of HWO. There is a strong push from within the community to develop a vacuum testbed in Europe to decrease these technology development cycles. This has also been discussed during the previous RD-HCI I workshop in the context of European infrastructure \citep{2025Ap&SS.370...29L}. Now is the time to start the actual development.\vspace{0.1cm}

\section{Ground-based facilities}
% Olivier Absil
In the context of HWO, ground-based facilities have three main purposes: (i) test and validate technologies and system-level concepts in a representative laboratory environment, (ii) develop observing strategies and post-processing methods on sky, and (iii) prepare and support the scientific exploitation of the mission. The first item was an important subject of discussion in WP1 (wavefront control), where it was stressed that a vacuum testbed would be required for Europe to make a significant contribution to this topic. Further synergistic technology development is showcased by the demonstration at the VLTI of interferometric techniques relevant for LIFE -- such as nulling, photonic chips, and metrology/fringe tracking systems. This aspect will not be further considered here; instead, we focus on the role of European ground-based telescopes.\vspace{0.1cm}

\textbf{Planetary Camera and Spectrograph synergy}. Developing observing strategies and post-processing methods relevant to HWO would greatly benefit from achieving contrast regimes significantly deeper than those currently attainable on 10-m class telescopes equipped with state-of-the-art high-contrast imaging instruments. In Europe, the Planetary Camera and Spectrograph (PCS; \citealt{2021Msngr.182...38K}) for the ELT offers an opportunity to advance these strategies and concepts at HWO-relevant contrast levels. However, the anticipated PCS timeline, with installation in the late 2030s at best, precludes its use in driving system-level decisions for HWO. Nevertheless, PCS development could still inform HWO in several ways, which merit further exploration. These include novel approaches to wavefront sensing, system-level optimisation of wavefront control, and improved exploitation of spectral information for both instrument control and scientific analysis. Beyond PCS, certain system-level concepts (e.g., wavefront control and observing strategies) could still be tested on-sky with 10-m class telescopes, albeit with limited relevance to HWO in terms of contrast.\vspace{0.1cm}

\textbf{Extremely Large Telescope preparatory programs}. The most significant contribution of European ground-based facilities to HWO is expected to lie in science preparation and follow-up. While a comprehensive review of science gaps and Europe’s potential role is beyond the scope of this section, several key contributions can be highlighted. Chief among these is early access to the Extremely Large Telescope and its first-generation instruments. MICADO and METIS, anticipated to achieve first light around 2030, followed by HARMONI and ANDES, will deliver new capabilities to advance our understanding of exoplanet atmospheres and demographics. These observations will provide essential context for exploring potentially habitable worlds, particularly by extending exoplanet spectroscopy into the super-Earth regime, one of HWO’s primary science gaps.\vspace{0.1cm}

\textbf{Target selection through high-precision radial velocity surveys}. Another major European asset is access to high-precision radial velocity instruments such as HARPS \citep{mayoretal2003}, HARPS-North \citep{cosentinoetal2012}, NIRPS \citep{wildietal2017}, CARMENES \citep{quirrembachetal2014} and ESPRESSO \citep{pepeetal2010}. These facilities can assist HWO target selection by identifying larger planets within the habitable zones of candidate stars and will be critical for follow-up observations. They will soon be complemented by dedicated facilities such as the Second Earth Spectrograph \citep[MPG/ESO 2.2m Telescope,][]{sturmeretal2024} and the Terra Hunting Experiment \citep[Isaac Newton Telescope,][]{Hall2018}. Mid-infrared instruments like METIS on the ELT \citep{Brandl2021}, and the ASGARD/NOTT nulling interferometer for the VLTI \citep{laugieretal2023}, could also identify stars with substantial habitable-zone dust, which would hinder direct imaging of terrestrial planets. A survey of all southern HWO targets with ASGARD/NOTT appears feasible within the next few years. More broadly, ELT instruments will be key in characterising the global architecture of HWO planetary systems through high-contrast imaging, and in determining which architectures favour the formation of truly habitable exoplanets.

\section{Coronagraphs}
% Axel Potier

The discussion focused on the future of coronagraph development across three wavelength regimes — visible (VIS), near-infrared (NIR), and ultraviolet (UV). Each regime presents distinct challenges and opportunities \citep{Chen2024}: VIS benefits from a strong heritage, NIR carries significant technical gaps but compelling science cases, and UV remains largely unexplored but strategically promising.  \vspace{0.1cm}

\textbf{Visible Coronagraphs}. VIS coronagraphy is the most mature, with extensive heritage from the JPL/High Contrast Imaging Testbed and Roman testbeds \citep{Seo2019,Mennesson2024}. The main challenge lies not in concepts, but in realistic modeling and tolerancing \citep{Zhou2020}. Alignment errors, such as deformable mirror misclocking, actuator offsets, and mask orientation, must be included in performance predictions. Current coronagraph architectures, such as shaped-pupil and vortex designs, remain dominant  \citep{Belikov2024} but early efforts on 2-3 architectures (e.g., one per national space agency) would provide additional alternatives. \vspace{0.1cm}

\textbf{Near-Infrared Coronagraphs}. NIR coronagraphy is considered scientifically essential \citep{Damiano2022}, and it is also the regime where synergies with ground-based efforts are strongest. HWO’s NIR goals align with the capabilities expected from the Planetary Camera and Spectrograph (PCS, \cite{Kasper2021}), which aims to achieve contrasts on the order of $10^{-9}$ in the NIR. This creates a valuable cross-over between space and ground, enabling technology synergies. However, applying VIS-style coronagraph designs in the NIR would cause the number of accessible targets to drop sharply at longer wavelengths \citep{Morgan2024}, while there is still a major gap for achieving small inner working angles around 60 mas ($\sim1\lambda/D$).  Still, the scientific case for NIR is compelling, and its overlap with PCS provides a strong argument for prioritizing development in this regime despite the risks.\vspace{0.1cm}

\textbf{Ultraviolet Coronagraphs}. 
UV coronagraphy is the least developed, but it offers unique potential for Europe's involvment in HWO. Unlike VIS and NIR, there is no significant coronagraph testbed experience in UV to date \citep{VanGorkom2025}. It is generally considered a separate instrument rather than a channel within the main coronagraph, which provides a unique opportunity for the European high-contrast imaging community to be involved in HWO. However, it may ultimately serve as a second-generation capability for HWO. A standalone ESA far-UV pathfinder mission, targeting moderate contrasts, with a modest 1–2 meter aperture could help fill an important exploration gap. A mission in this regime would build technical expertise and serve as a stepping stone toward future concepts such as starshades or more advanced instruments.\vspace{0.1cm}

%\textbf{Starshades}. 

Overall, the discussion highlighted how VIS coronagraphs provide a proven foundation, while NIR offers the strongest near-term scientific return through synergy with PCS, though significant technical gaps remain. UV, by contrast, represents a longer-term opportunity to expand into unexplored territory. Participants also debated the value of a dedicated European vacuum testbed, with differing views on whether it would enable crucial component testing or divert resources from instruments. Early strategic investment and European coordination between laboratories, space national agencies and ESA will be essential to balance these regimes and ensure coronagraph development advances effectively.

\section{Post-processing}
% Herve Le Corroler, Elisabeth Matthews

%This session addressed the post-processing techniques necessary for HWO. Note that further details are available in the white papers produced by the NASA Coronagraph Concept of Operations and Post-Processing (COPP) working groups \citep{LeCoroller2025}. 
We recalled that detecting Earth-like planets around nearby stars will involve operating in a new, previously unexplored contrast regime. Achieving a noise floor contrast of $3\cdot10^{-12}$ at 1-$\sigma$ is essential. At this contrast level, the effectiveness of classical Angular Differential Imaging (ADI) and Reference Differential Imaging (RDI) techniques is uncertain. RDI may be sensitive to line of sight jitter, and stellar diameter mismatches between the science target and the reference with a limit of $\approx 1$ mas mismatch \citep{Juanola-Parramon_2022JATIS...8c4001J}. Both ADI and RDI could be affected by star spots (on the science and/or reference star) and exozodiacal emission \citep{kammereretal2022AJ....164..235K, currieetal2023AJ....166..197C}. Additionally, RDI and ADI are sensitive to wavefront drifts of 10-50 pm RMS during slews, rolls, and observations \citep{Juanola-Parramon_2022JATIS...8c4001J}. Several proposed solutions have been presented:
\begin{itemize}
\item	DM and Angular speckle smoothing : both tested in lab on the NASA Decadal Survey Testbed (Credit: Stark, C., Roser, J.P.).
\item	WFS-based PSF : A calibration with telemetry acquired during the Science target observations, so no more need for RDI or ADI \citep{guyon_2022SPIE12185E..0EG}
\item	Keplerian algorithms: The goal is to use multi-epoch observation making use of short planets orbital motion to jointly increase the detection confidence and constrain the orbital parameters. %The Keplerian algorithms can be applied in conjunction with any high contrast imaging reduction that will be finally used (ADI, RDI, but also WFS-PSF, CDI, PDI, etc.).
\end{itemize}

To date, three Keplerian algorithms have been developed:
\begin{itemize}
\item K-Stacker \citep{lecoroller_2015tyge.conf...59L, nowak_2018A&A...615A.144N,lecoroller2020A&A...639A.113L,lecoroller_2024sf2a.conf..363L} is a brute-force approach. Several science results have already been obtained \citep{Wagner_2021NatCo..12..922W,lecoroller_2022A&A...667A.142L,tschudi_2024A&A...687A..74T}.

%with this algorithm. For example, on alpha Cen A VISIR observations \citep{Wagner_2021NatCo..12..922W}, K-Stacker has confirmed the detection of a C1 candidate in habitable zone of the star  \citep{lecoroller_2022A&A...667A.142L}. One of the highest contrasts ever achieved from the ground in high-contrast imaging ($10^{-8}$) was obtained using SPHERE-ZIMPOL with the K-Stacker algorithm around $\epsilon$ Eri \citep{tschudi_2024A&A...687A..74T}. The analysis with K-Stacker also yielded possible orbital parameters for $\epsilon$ Eri b, which are consistent with the solution proposed by \citep{Llop-Sayson_2021AJ....162..181L}.
\item	OCTOFITTER \citep{thompson_william_2023AJ....166..164T} uses a Markov-chain Monte Carlo (MCMC) within a Bayesian framework \citep{ruffio_2018AJ....156..196R}
\item	PACOME \citep{dallant2023AA679A38D, 2022dallant_GRETSI, 2022dallant,Dallant2024} integrates an exploration of the plausible orbits of the sought-for objects within an end-to-end statistical detection and estimation formalism. %The latter is extended to a multi-epoch combination of the maximum likelihood framework of PACO, which is a post-processing algorithm of single-epoch observations. %From this, PACOME derive a reliable multi-epoch detection criterion, interpretable both in terms of probability of detection and of false alarm and produce a plausible estimates of the orbital elements of the detected sources. 
\end{itemize}

MCMC Keplerian algorithms such as Octofitter or K-Stacker-MCMC \citep{LeCoroller2025} can combine independent datasets (e.g., Radial Velocity, Astrometry, High Contrast Imaging) to compute robust orbital parameters and planet flux. Theoretically, they can detect planets undetectable in a single dataset (e.g., HCI) by leveraging all available information. These methods are highly promising for combining multi-epoch observations and multiple techniques on HWO targets.

The brightness phase curve must be considered in the multi-epoch observational strategy. In combination with other techniques (RV and astrometry), observations can be scheduled to enhance planet detection chances.  
We discussed the need to compare an observing scenario with a single visit per target with multiple visits per target (see \citealt{LeCoroller2025}). The importance of using K-Stacker algorithms grows if we aim to determine orbital parameters. These parameters are crucial for better understanding planet formation \citep{Raymond2018MNRAS.479L..81R, Morbidelli2016IAUFM..29A...3M}. Additionally, it is vital to determine the phase period during observations to better constrain the atmosphere \citep[e.g., the relationship between the phase curve and clouds, see][]{nayaketal2017, damianoetal2020, carriongonzalezetal2021b, salvadoretal2024}. We discussed the confusion effect that can occur when multi-epoch recombination algorithms mistakenly combine multiple Keplerian objects (e.g., planet b with planet c) or a planet with background stars or bright speckles. According to \cite{Pogorelyuk_2022ApJ...937...66P}, using four epochs is generally sufficient to avoid this confusion effect. MCMC techniques like Octofitter or PACOME are less sensitive to this issue because they better statistically model noise, PSF, and flux response.\vspace{0.1cm}

A polarization mode could aid in detecting life signatures and reducing false alarm rates \citep{gordon2025ApJ...983..168G}. Additionally, it may help resolve the degeneracy between radius and albedo \citep{Roccetti2025A&A...702A.262R}. We also discussed the extent to which we can understand polarization if precursor observations do not utilize it. We recommended using the polarization mode with the Roman Space Telescope that will serve as a testbed for HWO.\vspace{0.1cm} 

We propose that during the next HCI R\&D Europe meeting, discussions and presentations on instrumentation (including WFS, telemetry, telescope, etc.) and data reduction algorithms be conducted in the same room.\vspace{0.1cm}

%\textbf{Conclusion \& follow-up actions}. 

%Beyond previous considerations,  
%The reduction algorithms must be developed concurrently with instrumentation to achieve a noise floor of $10^{-12}$. Several new reduction techniques were presented, where the instrumentation is modified for these methods: A technique that records Wavefront sensor telemetry to compute a reference PSF \citep{guyon_2022SPIE12185E..0EG}; a technique that uses the Deformable Mirror (DM) to smooth speckles (tested in the lab on the NASA Decadal Survey Testbed. Credit: Stark, C., Roser, J.P.); a technique with metrology at the curvature center of the primary mirror to control the wavefront within a few picometers, potentially allowing the replacement of ADI and RDI with ODI for the nearest stars within 10 parsecs \citep{LeCoroller2025}. A polarization mode could aid in detecting life signatures and reducing false alarm rates \citep{gordon2025ApJ...983..168G}. Additionally, it may help resolve the degeneracy between radius and albedo \citep{Roccetti2025A&A...702A.262R}. We also discussed the extent to which we can understand polarization if precursor observations do not utilize it. We recommended using the polarization mode with the Roman Space Telescope that will serve as a testbed for HWO. We propose that during the next HCI R\&D Europe meeting, discussions and presentations on instrumentation (including WFS, telemetry, telescope, etc.) and data reduction algorithms be conducted in the same room.\\

Future ``data challenges'' will require the simulation of observing scenarios employing multi-epoch and multi-technique strategies. Observations will be divided into several epochs (e.g., four) to constrain the orbital parameters within the minimum total observing time necessary for exoplanet detection. Data reduction will be performed using Keplerian-based algorithms such as K-Stacker, Octofitter, and PACOME. The impact of planetary brightness phase variations in reflected light, as well as confusion effects, must be thoroughly investigated to optimize the scheduling of observations and the application of multi-epoch and multi-technics algorithms.
For all these studied, exo-Earth atmospheres and spectrum at various resolutions will have to be simulated. \vspace{0.1cm}

%\textbf{Finalized List of Actions:}
%\begin{itemize}
%    \item Compile a comprehensive list of existing tools for simulating spectra of exo-Earths and phase curves, along with relevant references and papers (useful for HWO and LIFE). Relevant European expertise exists in this area of atmospheric modelling \citep[e.g.][]{garciamunoz2015IJAsB..14..379G, roccettietal2025A&A...700A..62R}
%    \item Coordinate with the NASA COPP team to align efforts on data simulation activities.
%    \item Organize and execute a data challenge to evaluate reduction algorithms using statistical approaches (e.g., true/false positives, ROC curves) and test various observing scenarios (e.g., single observation per science target vs. multi-epoch strategies).
%    \item Identify and define areas of collaboration between the LIFE and HWO teams, such as simulations of Earth-like planet spectra and compilation of target lists.
%\end{itemize}

\section{Nulling}
% Roman Laugier, Denis Defrere
For nearby habitable zone planets, resolving the planet light at wavelength $\geq 1.5 \mathrm{\mu m}$ pushes the limit of  coronagraphs on 8m-class telescopes. Spectral characterization using the thermal emission ($\approx 4-19\mu m$), would reveal the greenhouse effect in action, and components such as $\mathrm{CH_4}$ and $\mathrm{O_3}$. This suggests a dimension $\approx 10\times$ larger, and interferometry is the only plausible option. The detector aspects of the problem are discussed in Section \ref{sec:dectectors}.\vspace{0.1cm}

\textbf{Formation flying}. The formation flying aspect of space interferometry is sometimes portrayed as unsurmountable. However, the landscape has changed since the stop of DARWIN \citep{Vasile2021} with the missions PRISMA \citep{Bodin2012a}, PROBA-3 \citep{Jorgensen2023, Capobianco2021}, the development of LISA, including LISA Pathfinder \citep{Armano2024, Armano2024a}, GRACE-FO \citep{Murbock2023}. This has demonstrated the feasibility of formation flight to millimeter precision, from which the control of the optical path can be handed off to short but highly precise internal delay lines. There are also a number of ongoing and proposed missions to demonstrate the stability of formation flying in the context of interferometry during beam transport and fringe tracking, including the ground-based \textit{Pyxis} interferometer \citep{Hansen2023}, and the space-based STARI \citep{Monnier2024} and SEIRIOS \citep{Matsuo2022} missions. \vspace{0.1cm}

\textbf{High Sensitivity}. Demonstrating high sensitivity in mid-infrared observations is a key requirement for proving the feasibility of missions targeting rocky exoplanets. Due to the low flux from these sources, and the high thermal background from instrumentation at mid-infrared wavelengths, such observations must be made at cryogenic temperatures below 50~K. This imposes many optomechanical challenges in combining these low temperatures with the high precision needed for nulling interferometry. This objective, combining high contrast with high sensitivity, has been endorsed by ETH Zürich through the NICE (Nulling Interferometry Cryogenic Experiment) project, which aims to test a cryogenic nulling beam combiner operating at high contrast and flux levels comparable to realistic scenarios. The experimental setup is described in \cite{Ranganathan2022,ranganathan2024,Birbacher2026}. It is also worth mentioning that Europe has substantial heritage in developing highly sensitive mid-infrared missions at low temperatures, such as for JWST/MIRI, ELT/METIS and VLTI/MATISSE.\vspace{0.1cm}

\textbf{Broad Bandwidth}. A related challenge is achieving nulling performance across a broad spectral range, typically from $4 \,\mathrm{\mu m}$ to $18.5 \,\mathrm{\mu m}$ as required for LIFE \citep{quanzetal2022}. This requires precise control of phase, amplitude, and polarization across the entire bandwidth. Minimizing the number of spectral splits is critical to reduce system size, mass, and complexity. A promising approach is the adaptive nuller proposed by \cite{Lay2003}, which uses a dispersed delay line combined with a deformable mirror. The beam is dispersed by a spectrograph (and optionally split by a Wollaston prism), enabling independent control of wavelength and polarization channels before recombination. Excellent performance was demonstrated with this concept \citep{Peters2010}, but the design process was not fully documented, limiting reproducibility at TRL5. The aforementioned cryogenic requirement also requires such a deformable mirror to work sufficiently at sub 50~K conditions, which while previously demonstrated \citep{Takahashi2017}, requires active development to rapidly increase its TRL. Further discussion on DMs are detailed in the following section. These problems are actively being worked on by multiple European groups such as at ETH Zurich \citep{Birbacher2026} and TU Delft 
\citep{Loicq2024}, investigating both the adaptive nuller as well as alternatives not requiring cryogenic deformable mirrors.
\vspace{0.1cm}

\textbf{Spatial Filtering}. Spatial filtering remains a challenge for broadband scaling. Such devices need to remain single mode, have low losses and allow for high coupling efficiency over the entirety of the bandpass (or substantial parts thereof). One promising approach, outlined by \cite{Ireland2024}, combines specialized injection optics with photonic crystal fibers. This concept is currently at TRL2, with a roadmap to TRL4 recently funded.\vspace{0.1cm}

\textbf{Vibration Mitigation}. Technologies for measuring and mitigating structural vibrations are essential. LIFE team members have led improvements to the VLTI vibration monitoring system \citep{Laugier2024} for projects such as GRAVITY+ \citep{gravityplus2025arXiv250921431G} and the upcoming Asgard/NOTT nuller \citep{Laugier2023,Defrere2024}. Current strategies combine accelerometers for high-frequency vibrations with laser metrology for low-frequency components \citep{Birbacher2022,Birbacher2024}.\vspace{0.1cm}

\textbf{Integrated Optics}. Integrated optics offers advantages for space interferometry due to its low mass and high stability. However, the challenges of operating over very broad mid-infrared bands make it difficult for nulling in the science band primarily due to immature and non-transmissive platforms. There are a number of efforts into quickly maturing this technology \citep[e.g.][]{MontesinosBallester2024}, but whether such efforts will be fruitful in the near-term remains to be seen. Its application for fringe tracking in the near-infrared (around $2.5 \,\mathrm{\mu m}$) is promising, as demonstrated by the GRAVITY beam combiner at TRL7.\vspace{0.1cm}

%Ground-based demonstrations of deep nulling interferometry \citep{lacouretal2019A&A...624A..99L} and exoplanet detection have highlighted the importance of supporting the development and science programs of ongoing mid-IR nullers such as LBTI \citep{Ertel2022}, Asgard/NOTT \citep{Defrere2024}, or the Planet detection testbed \citep{martinetal2012ApOpt..51.3907M}.

\section{Deformable mirrors}
% Markus Kasper

During the DM breakout session, we reviewed and discussed the state-of-the-art in Deformable Mirrors (DMs) for Extreme Adaptive Optics (XAO), with a focus on European suppliers. \vspace{0.1cm}

\textbf{ALPAO-Bertin Developments}.  
ALPAO-Bertin (Grenoble, France) continues to advance its technology toward high actuator count, fast-response DMs suitable for XAO. Stefan Stroebele reported that the recently delivered MAVIS DMs, featuring 64 actuators across the aperture, exceeded stability requirements with fewer than five inactive actuators. \vspace{0.1cm}

\textbf{ESO XAO DM Prototype}.  
ESO is developing an XAO DM prototype to meet the requirements of the ELT Planetary Camera and Spectrograph (PCS). This is a 4–5 year development effort that began in early 2024. The first work package (WP1: preliminary design) is nearing completion, and ESO has exercised options for WP2 (final design) and WP3 (optical head Manufacturing, Assembly, Integration, and Testing - MAIT). Funding opportunities for WP4 (prototype MAIT) and WP5 (prototype ownership) are being explored.  
Key requirements include:
\begin{itemize}
    \item At least 128 actuators across the pupil
    \item Small stroke settling time $<300~\mu$s
    \item $\pm3~\mu m$ stroke at 0.2~nm resolution
    \item Integrated drive electronics within the DM housing
\end{itemize}\vspace{0.1cm}

\textbf{Multi-Stepping Actuation Concept}.  
Fast temporal response may be achieved through a multi-stepping actuation concept currently under development by ALPAO-Bertin.\vspace{0.1cm}

\textbf{Thermal-Optical Spatial Light Modulator (T-SLM)}.  
Jonas Kühn reported on his collaboration with ALPAO-Bertin on the thermal-optical SLM (T-SLM) under the ``RACE-GO'' ERC grant. The T-SLM uses spherical gold nanoparticles embedded in a polymerized PDMS layer between two sapphire windows. An acousto-optic driven pulsed illumination laser distributes temperature in 2-D within the T-SLM, enabling polarization-insensitive phase shifts. A first prototype is expected soon, meeting key requirements:
\begin{itemize}
    \item Stroke $>1.63~\mu m$
    \item Transmission $>$50\%
    \item 500$\times$500 resolution elements with $20–50~\mu m$ pitch
    \item Response time $<$10~ms
\end{itemize}\vspace{0.1cm}

\textbf{Fraunhofer IPMS Developments}.  
The Fraunhofer-Institut für Photonische Mikrosysteme (IPMS, Dresden, Germany) develops spatial light modulators using micromirror arrays on semiconductor chips. Sebastiaan Haffert reported on negotiations to modify (bin) the front structure of their segmented piston device to achieve larger resolution elements with higher fill factor and piston + tip-tilt actuation. This modification would be relatively inexpensive and enable high actuator counts, though stroke would be limited to $<1~\mu m$, making it suitable for third-stage XAO or Non Common Path Aberrations (NCPA) correction. Stability and print-through remain to be evaluated.\vspace{0.1cm}

\textbf{NASA DM Roadmap}.  
Karl Stapelfeld presented NASA’s DM roadmap developed under the Exoplanet Exploration Program (ExEP). Three potential suppliers were identified for HWO DMs: AOA Xinetics, Boston Micromachines, and ALPAO-Bertin. NASA plans to provide seed funding to these vendors to develop credible fabrication plans for DMs meeting preliminary requirements:
\begin{itemize}
    \item 96 actuators across the aperture
    \item 500~nm PV stroke
    \item 1~nm RMS flatness
    \item Actuator pitch $<$1~mm
    \item 100\% actuator yield
    \item Stability $\sim$5~pm RMS/hr and resolution 2~pm
\end{itemize}
A dedicated NASA facility will be required to test and characterize large-format DMs to HWO performance levels.

\section{Telescope Design \& Optics}
% Oliver Krause
Defining the telescope architecture, and the requirements of its mechanical and optical components is a key to identify potential European contributions and associated partners. The rapid developments around the establishment of the HWO project office and the work packages motivated the inclusion of this topic on the agenda of this RD-HCI II workshop, which had not been covered in the RD-HCI I.\vspace{0.1cm}

\textbf{Identified Priorities and European Expertise}.
The group discussed several areas where Europe has world-leading technological expertise and could contribute to the HWO mission. Possible avenues for these contributions include ESA-managed packages or component-level involvement through national agencies.\vspace{0.1cm}

\textbf{Mirrors}. With lessons learned from JWST and the deployment of adaptive controls as standardly done on the ground \citep{Pueyo2022}, we now know that HWO can be built with a segmented primary mirror while still providing the required wavefront error limits and stability, and the ability to fold into future launchers. Although NASA is expected to lead this effort, Europe has significant expertise in this area and could contribute mirror segments or the secondary mirror through ESA.\vspace{0.1cm}

\textbf{Ultra-Stable Structures and Ultra-Precision Optics}.  
Accuracy requirements comparable to space-based HCI, covering stability, fabrication, verification, and metrology down to single-digit picometer levels, exist in the domain of EUV semiconductor optics, where Europe is a global leader. European industry (e.g., CARL ZEISS SMT) could be key partners in developing and fabricating critical optical components for an HWO coronagraph. European efforts in space-based reconnaissance will also contribute to this technology R\&D.\vspace{0.1cm}

\textbf{High-Performance Optical and Opto-Mechanical Components}.  
Specialized components such as dichroics, bandpass filters with very high rejection, low-scintillation optics, and low-power, high-precision mechanisms will be required for a space-based coronagraph. For a U.S.-led mission, these could represent smaller-scale European contributions at the instrument level.\vspace{0.1cm}

\textbf{Occulter/Starshade}.  
Discussion also included starshade development which is moving from concept toward mission readiness. The NASA’s Exoplanet Exploration Program has spent the last several years maturing the core starshade technologies, ultra-deep starlight suppression, precise formation sensing and control over tens of thousands of kilometers, and reliable deployment of very large, petal-shaped structures, under the “Starshade to TRL-5 (S5)” effort. This effort includes lab tests, scaled experiments, and detailed modeling showing the required contrast is achievable\footnote{see NASA Starshade Closeout Briefing Report:
Technology Status and Applicability to the
Habitable Worlds Observatory, https://science.nasa.gov/astrophysics/programs/exep/technology/starshade/}. At the same time, Europe leads in key technologies such as electric propulsion and formation-flying (e.g., ESA PROBA-3 mission). Another potential contribution is the starshade receiver instrument, which could occupy one of the five HWO instrument bays and provide science data. These represent high-profile opportunities for European (ESA) involvement. \vspace{0.1cm}

\textbf{Identified Gaps and Next Steps} 
The group discussed strategies to bring these ideas to the attention of the ESA Science Programme Committee (SPC) and NASA/ESA inter-agency discussions. Traditionally, such proposals are raised at the national agency level through SPC delegates. Additional discussion points included:
\begin{itemize}
    \item The concept of a small unobscured off-axis precursor space telescope dedicated to HCI, discussed at RD-HCI I, has been retired due to the lack of a compelling science case.
    \item Telescope compatibility for polarimetry will be addressed in HWO design studies, subject to science requirements.
    \item Regarding additional coronagraph channels, NIR is prioritized over UV.
\end{itemize}

The group concluded that, following the ESA HWO townhall in May 2025, dialogue should begin with SPC members and within the ESA SPC. Industry partners should also be engaged to initiate discussions on the topics and potential contributions outlined above.

\section{Detectors} \label{sec:dectectors}
% Daniel Dicken, Pieter de Visser
The discussion focused on detectors for future space-based high-contrast imaging missions, particularly HWO and LIFE.

\textbf{Current Development Landscape}.  
Detector development for HWO is currently led by industry, with CCDs for UV/visible wavelengths and Avalanche Photodiodes (APDs) for near-IR being pursued by companies such as \textbf{Teledyne} and \textbf{Leonardo} \citep{Pratlong2024,Baker2024}. This industrial focus presents challenges for defining clear European contributions, which may be limited to system-level testing or integration. However, a French team working on an astrometry instrument for HWO \citep{malbetetal2025arXiv251018920M, amiauxetal2025arXiv251107113A} is collaborating with \textbf{Pyxalis}, which is providing large-format CMOS detectors for evaluation, a potential European contribution to the high-resolution camera.\vspace{0.1cm}

\textbf{CCD and Photon-Counting Technologies}.  
Current Electron Multiplying CCD (EMCCD) technologies do not meet HWO’s stringent requirements for single-photon counting. Present detectors achieve dark noise levels of $\sim$1.7~e\textsuperscript{-} RMS, whereas HWO requires $<$0.25~e\textsuperscript{-} RMS. Teledyne’s CIS303 and CIS304 detectors already satisfy most specifications for the high-resolution camera, demonstrating high TRL. A promising pathway to photon-counting CCDs is \textbf{Skipper mode}, which enables multiple non-destructive readouts per pixel, significantly reducing readout noise and allowing single-photon detection.\vspace{0.1cm}

\textbf{Linear-Mode APDs (LmAPDs)}.  
LmAPDs based on HgCdTe arrays are designed for ultra-low flux NIR astronomy, detecting signals down to a few photons per pixel per hour. A collaboration between the Institute for Astronomy (University of Hawaii) and Leonardo UK aims to mature this technology for HWO \citep{Claveau2024}. Current testing involves 1k$\times$1k arrays, with plans to scale to 2k$\times$2k. Recent progress includes dark currents reduced to 0.07~e\textsuperscript{-}/pixel/kilosecond, approaching negligible levels. Read noise remains the main barrier, but LmAPDs mitigate this by amplifying photon signals before readout, positioning them as strong candidates for “noise-free” photon counting.\vspace{0.1cm}

\textbf{Energy-Resolving Detectors}.  
Energy-resolving detectors were discussed as transformative for science yield, requiring new approaches to data acquisition and mission design \citep{Howe2024,Steiger2024}. These detectors need significant maturation in spectral resolution ($R$), efficiency, yield, and radiation hardness. Achieving TRL5 and demonstrating performance at scale are key milestones.\vspace{0.1cm}

\textbf{MKIDs (Microwave Kinetic Inductance Detectors)}.  
MKIDs offer noiseless detection and could improve science yield by $>$30\%, but require trade-off studies. They have limited spectral resolution ($R\sim100$), need radiation testing, and require ultra-low-temperature cooling (100--300~mK). European expertise exists within SRON, Lux Lab, and the Paris Observatory, with SRON expected to gain operational experience via the PRIMA mission. While MKIDs remain promising, their inclusion in HWO’s first-generation instruments is uncertain due to passive cooling and strict vibration limits in the current mission design.\vspace{0.1cm}

\textbf{Detector Challenges for LIFE}.  
For LIFE, achieving sufficient sensitivity across 4--18.5~$\mu$m is critical. Historically, Si:As and Si:Sb detectors (e.g., MIRI) were suitable, but production ceased following Raytheon’s retirement of expertise. HgCdTe detectors are mature for 1--5~$\mu$m but remain experimental at longer wavelengths, with high noise levels (e.g., GEOSNAP). Teledyne’s HgCdTe GEOSNAP detectors (6–10.5~$\mu$m cutoff) have been selected for NEO Surveyor \citep{Zengilowski2022}. ESA’s plans for European-made HgCdTe detectors remain unclear, and a community-driven effort may be needed to revive Si:As or extend HgCdTe capabilities.\vspace{0.1cm}

\textbf{Emerging Alternatives for LIFE}.  
MKIDs and SNSPDs (Superconducting Nanowire Single-Photon Detectors) are being explored for mid-IR photon counting. SRON and JPL are developing MKIDs for PRIMA \citep{Baselmans2022,Day2024}, with sensitivity demonstrated up to 24~$\mu$m. SNSPDs developed by JPL and NIST \citep{Taylor2023,Hampel2024} show photon-counting capability in the 10--29~$\mu$m range. Both technologies require significant improvements in efficiency, filtering, and radiation hardness, as well as ultra-low-temperature cooling (100--300~mK), which will impact mission design.\vspace{0.1cm}

\textbf{Next Steps}.  
It was agreed that ESA’s plans for European-made MCT detectors should be investigated before the next meeting.

\section{Conclusions} \label{sec:conclusions}

From the three days of presentation and discussion at the European R\&D for Space-based HCI - II workshop, the key following conclusions and actions emerged:

\begin{itemize}

\item Europe has strong expertise in adaptive optics and WFSC, but most developments have been for ground-based systems, requiring significant adaptation for space use. Advancements are needed in several areas: developing non-linear and machine-learning-based control algorithms, optimizing electric field sensing beyond existing PWP and SCC methods, and adapting real-time computing systems to meet the stringent requirements of space environments. Europe also lacks dedicated vacuum testbeds for high-contrast imaging, hindering end-to-end verification and slowing innovation. To remain competitive and reduce development cycles, \textbf{the community strongly supports establishing a European vacuum testbed and expanding space-qualified WFSC infrastructure.}\vspace{0.1cm}

\item For future coronagraph developments for which the European community could play a key role, \textbf{the NIR coronagraph represents a clear opportunity}. It is scientifically vital and benefits from strong synergies with ground-based systems like PCS, though major technical gaps remain, particularly in achieving small inner working angles. \textbf{UV coronagraphy, though least developed, presents also a unique opportunity for European leadership}, with potential for a dedicated ESA pathfinder mission to build expertise and bridge current capability gaps.\vspace{0.1cm} %\textbf{Coordinated European investment and collaboration among laboratories, national agencies, and ESA will be essential to balance these efforts and advance coronagraph technology effectively.}

\item \textbf{For ground-based facilities, Europe will play a crucial role in supporting the HWO and LIFE mission through technology validation, development of observing strategies, and scientific preparation and follow-up.} They will enable testing of wavefront control concepts and advanced post-processing techniques, as well as provide key exoplanet spectroscopy and characterization toward the super-Earth regime. They will also demonstrate interferometric techniques at high contrast/sensitivity. Europe’s high-precision radial velocity instruments and mid-infrared facilities will also support target selection, follow-up, and detection of habitable-zone dust.\vspace{0.1cm}
%- technical conclusion of the technical sections; which aspects emerged (e.g. NIR path on HWO coro instrument)\\

\item \textbf{Data reduction algorithms must be developed alongside instrumentation.} Several advanced techniques are being explored, including using wavefront sensor telemetry for reference PSFs, deformable mirrors to smooth speckles, and picometer-level wavefront control with primary mirror metrology. Polarization modes may help detect life signatures, reduce false positives, and resolve radius-albedo degeneracies, with precursor observations on the Roman Space Telescope recommended. \textbf{Future work includes simulating multi-epoch and multi-technique observing scenarios, applying Keplerian-based reduction algorithms, and modeling exo-Earth atmospheres and phase curves}.\vspace{0.1cm} %Key actions involve cataloging existing simulation tools, coordinating with NASA for data simulations, organizing data challenges to test reduction algorithms statistically, and identifying collaborative work between LIFE and HWO teams to optimize exoplanet detection and characterization.

\item For nulling technologies, observing nearby habitable-zone planets at wavelengths beyond near-IR and characterizing their thermal emission at mid-IR requires interferometry. \textbf{Key technical challenges include achieving high sensitivity, broad-band nulling with precise control of phase, amplitude, and polarization, and developing spatial filtering solutions.} Adaptive nullers and photonic crystal fiber spatial filters provide promising approaches. Structural vibration mitigation, combining accelerometers and laser metrology, will be also critical. Ground-based facilities will be essential for demonstrating deep nulling and exoplanet detection, supporting technology development for future space interferometry missions. Europe is the primary contributor to many of these technological challenges, and has substantial heritage through the participation in the LBTI, VLTI and LIFE projects. \vspace{0.1cm}

\item Deformable mirror (DM) developments for extreme adaptive optics (XAO) include high-actuator-count, fast-response systems, the ELT XAO DM prototype with stringent settling time and stroke requirements, as well as emerging concepts such as multi-stepping actuation, thermal-optical spatial light modulators, and micromirror arrays. NASA’s DM roadmap for HWO targets very high stability, fine resolution, and full actuator yield, while identifying key suppliers. In this context, Europe is well positioned to contribute, with a strong ecosystem of industrial and research actors. Developments from ALPAO-Bertin, ESO, and novel SLM-based approaches already align closely with HWO requirements in stability, actuator density, and system integration. \vspace{0.1cm}

\item The RD-HCI II workshop addressed telescope architecture for HWO, focusing on areas where Europe could contribute through ESA-managed packages or national initiatives. Key priorities include mirror designs, ultra-stable structures, high-performance opto-mechanical components and precision optics leveraging Europe’s expertise in semiconductor and space technologies. Additional opportunities lie in high-performance optical components and starshade technologies, where Europe leads in propulsion and formation flying. \textbf{The group emphasized the need for early engagement with ESA’s Science Programme Committee (SPC) and industry partners to explore these contributions.} Design considerations such as polarimetry compatibility and prioritization of NIR coronagraph channels over UV were noted, while the concept of a small precursor telescope was retired due to lack of a strong science case.\vspace{0.1cm}

\item \textbf{Detector development for HWO and LIFE is currently industry-led}, with CCDs for UV/visible and Avalanche Photodiodes (APDs) for NIR pursued by companies such as Teledyne and Leonardo. European contributions could emerge through collaborations, such as Pyxalis supplying large-format CMOS detectors for HWO’s high-resolution camera. Current EMCCDs do not meet HWO’s single-photon counting requirements, but Skipper-mode CCDs offer a promising path. Linear-mode APDs (HgCdTe) show progress toward “noise-free” photon counting for NIR, while energy-resolving detectors, MKIDs, and SNSPDs offer transformative potential but require significant maturation, low-temperature cooling (100–300\,mK), and trade-off studies for mission feasibility. For LIFE, mid-infrared sensitivity (4–18.5\,$\mu$m) is critical, but traditional detectors like Si:As and Si:Sb are no longer produced, and extending HgCdTe detectors to longer wavelengths remains challenging. European efforts, including SRON and other groups, are developing MKIDs and other superconducting photon-counting detectors, but efficiency, dark counts, and full-band performance require further work. \textbf{Coordination on European MCT detector plans and broader community development is essential to meet mission requirements.}
\end{itemize}\vspace{0.1cm}

\section{Next Steps} \label{sec:NextSteps}

\textbf{Overall, European communities are increasingly organizing on a national scale, and it is essential to extend this momentum to a broader European level.} While the first two workshops were held by invitation, the next phase aims to open participation to a wider group of stakeholders. The general conclusions so far confirm Europe’s strong interest in contributing to HWO/LIFE and in pursuing R\&D and HCI research, fields in which Europe holds significant potential regardless of external developments. There are clear technological opportunities to advance through higher TRL levels and address the HWO and LIFE gap lists that remain uncovered elsewhere. However, achieving this requires robust infrastructures and shared resources. Moving forward, the organization of upcoming activities should focus on maintaining engagement and motivation—avoiding long gaps between meetings, scheduling teleconferences every two months, and ensuring continuous progress tracking. Finally, this report should serves as a stepping stone to connect and consolidate the various efforts currently underway across European institutions and countries, providing an overview that strengthens collective coordination and collaboration.

%- Communities are starting to organize on a national scale, we need to continue this effort on a European level,\\
%- first two workshops were by invitation, seeking now to open it up to wider group\\
%- General conclusions confirming that Europe has clear interest in contributing to HWO/LIFE. Europe has clear interest in pursuing R\&D/HCI research (no matter what); Technological opportunities in Europe  to advance in TRL levels,  fill the HWO \& LIFE gap list (not covered elsewhere),\\
%- Need for infrastructures/resources\\
%- Concluding remarks on the organization of upcoming activities: Keep motivation with people, try to avoid meeting only in two years from now, maintain momentum, telecons every ~2 months or so, keep track of this collective initiative\\
%- this report is also there to sort of glue together the efforts happening in institutions and countries (some sort of summary or mention of those efforts)

\section{Actions} \label{sec:Actions}

\begin{enumerate}

    \item \textbf{Maintain Momentum}\vspace{0.1cm}
    \begin{enumerate}
        \item Schedule regular teleconferences (approximately every two months), keeping participants engaged, and tracking progress on this collective initiative. 
        \item Open the next workshops more broadly to the community (no restricted number of participants) and ensure greater participation from both the interferometric community and the LIFE community.
        \item Organise a follow up meeting in 2026? 
    \end{enumerate}    \vspace{0.3cm}
    
    \item \textbf{Establish a European vacuum testbed and expand space-qualified WFSC infrastructure}  \vspace{0.1cm}
    \begin{enumerate}
        \item Design a collaborative execution plan across the HCI community to realize these infrastructure goals.
        \item Coordinate to propose for an EU synergy grant for a European vacuum test bed - November 2026. 
    \end{enumerate}        \vspace{0.3cm}

    \item \textbf{Advance coronagraph technology through coordinated European investment}  \vspace{0.1cm}
    \begin{enumerate}
        \item Develop and mature NIR coronagraph capabilities, addressing technical gaps such as achieving small inner working angles.  
        \item Initiate UV coronagraph development and pursue a dedicated ESA pathfinder mission to build expertise and bridge capability gaps.  
        \item Foster collaboration among laboratories, national agencies, and ESA to balance efforts and accelerate progress.
    \end{enumerate}\vspace{0.3cm}

    \item \textbf{Leverage ground-based facilities to support HWO and LIFE missions}  \vspace{0.1cm}
    \begin{enumerate}
        \item Validate technologies, develop observing strategies, and prepare for scientific exploitation and follow-up using European observatories.
    \end{enumerate}        \vspace{0.3cm}

    \item \textbf{Post-processing} \vspace{0.1cm}
    \begin{enumerate}
        \item Compile a comprehensive list of existing tools for simulating spectra of exo-Earths and phase curves, along with relevant references and papers (useful for HWO and LIFE).
        \item Coordinate with the NASA COPP team to align efforts on data simulation activities.
        \item Organize and execute a data challenge to evaluate reduction algorithms using statistical approaches (e.g., true/false positives, ROC curves) and test various observing scenarios (e.g., single observation per science target vs. multi-epoch strategies).
        \item Identify and define areas of collaboration between the LIFE and HWO teams, such as simulations of Earth-like planet spectra and compilation of target lists.
    \end{enumerate}\vspace{0.3cm}

    \item \textbf{Nulling Technologies} \vspace{0.1cm}
    \begin{enumerate}
        \item  Pursue the development and testing of adaptive nullers and photonic crystal fiber spatial filters, implementing vibration mitigation strategies using accelerometers and laser metrology, and leveraging ground-based facilities to demonstrate deep nulling and exoplanet detection
    \end{enumerate}\vspace{0.3cm}

    % \item \textbf{Deformable Mirrors} \vspace{0.1cm}
    % \begin{enumerate}
    %     \item 
    % \end{enumerate}\vspace{0.3cm}

    \item \textbf{Telescope Design \& Optics}  \vspace{0.1cm}
    \begin{enumerate}
        \item The group concluded that, following the ESA HWO townhall in May 2025, dialogue should begin with SPC members and within the ESA SPC.
        \item Industry partners should also be engaged to initiate discussions on the topics and potential contributions outlined above.
    \end{enumerate}

\end{enumerate}

\bmhead{Acknowledgements}

Gael Chauvin would like to thank the Max Planck Institute for Astronomy, the Max Planck Society and the Haus der Astronomie for their support. Gael Chauvin would like to thank the Scientific Organizing Committee: Iva Laginja, LIRA (France), Pierre Baudoz, LIRA (France), David Doelman, SRON (Netherlands), Olivier Absil, Université de Liège (Belgium),  Oliver Krause, MPIA (Germany), Mamadou N'Diaye, Lagrange/OCA (France), David Mouillet, IPAG (France), Elodie Choquet, LAM (France), Daniel Dicken, STFC (UK), Michiel Min, SRON (NL), Sebastiaan Haffert, Leiden (NL), Oscar Carrion-Gonzalez (MPIA) and Gael Chauvin, MPIA (Germany); and the Local Organizing committee: Gael Chauvin, MPIA (Germany), Oscar Carrion-Gonzalez, MPIA (Germany), Macarena Vega Pallauta, MPIA (Germany), Matthieu Ravet, MPIA (Germany), and particularly Carola Jordan, MPIA (Germany) for their full support during before, during and after the workshop.

\section*{Declarations}

Not applicable
% - US community clearly highlighted the fact that the NIR coronagraph could be an important contribution for the European community\\
% - Interest in an intermediate mission in Europe with ESA before HWO/LIFE?\\
% - Vacuum testbed (EU synergy grant?)\\
% - discussion also about the starshade\\
% - Consistent structure of each section. Approx.: European expertise on the topic / Ongoing activities / Identified gaps / Identified next steps \\

% \bibliographystyle{science}
\bibliography{nulling}

\end{document}